\documentclass[10pt, a4paper, twocolumn]{IEEEtran}

\usepackage{amsmath,epsfig,amssymb,verbatim,amsopn,cite,subfigure,multirow}
\usepackage{algorithm,algorithmic}
\usepackage{balance,nopageno}
\usepackage{algorithm,algorithmic}
\usepackage[usenames,dvipsnames]{color}
\usepackage[all]{xy}  
\usepackage{url}
\usepackage{amsfonts}
\usepackage{amssymb}
\usepackage{epsfig}
\usepackage{epstopdf}
\usepackage{slashbox}
\usepackage{bm}
\usepackage{balance}
\usepackage[margin=17.0mm,top=20mm,bottom=18mm]{geometry}

\newcommand{\ve}[1]{\boldsymbol{#1}}
\newcommand{\E}[1]{\mathbb{E}\left\{#1\right\}}

 \newcommand{\vu}{\ve{u}}
 \newcommand{\vv}{\ve{v}}

\newcommand{\qf}{{\bf f}}

\newcommand{\qh}{{\bf h}}

\newcommand{\qn}{{\bf n}}

\newcommand{\qu}{{\bf u}}
\newcommand{\qv}{{\bf v}}
\newcommand{\qw}{{\bf w}}

\newcommand{\qA}{{\bf A}}
\newcommand{\qB}{{\bf B}}

\newcommand{\qE}{{\bf E}}

\newcommand{\qH}{{\bf H}}
\newcommand{\qI}{{\bf I}}

\newcommand{\qW}{{\bf W}}

\newcommand{\Snr}{\sigma_{R}^2}
\newcommand{\Snu}{\sigma_{n_{1}}^2}
\newcommand{\Snuu}{\sigma_{n_{2}}^2}
\newcommand{\Ps}{P_S}
\newcommand{\Pre}{P_R}

\newcommand{\Nrx}{{\mathsf{N_R}}}
\newcommand{\Ntx}{{\mathsf{N_T}}}
\newcommand{\SUu}{\text{CU}_{1}}

\newcommand{\SUuu}{\text{CU}_{2}}

\newcommand{\Sap}{\sigma^2_{\mathsf{SI}}}

\newcommand{\Ith}{I_\mathsf{th}}

\newcommand{\trac}{\mathsf{tr}}

\newcommand{\ThankFour}{The work of Z. Ding was supported by the UK EPSRC under grant number EP/N005597/1. It was also supported by the European Union H2020-MSCA-RISE-2015 under grant number 690750.}

\title{ Joint Beamforming Design and Power Allocation for Full-Duplex NOMA Cognitive Relay Systems}

\author{{Mohammadali Mohammadi$^\dag$, Batu K. Chalise$^*$, Azar Hakimi$^\dag$, Himal A. Suraweera$^\ddag$, and Zhiguo Ding$^\S$}\\
\small{$^\dag$Faculty of  Engineering, Shahrekord University, Iran\\
$^*$Department of Electrical Engineering and Computer Science, Cleveland State University, USA\\
$^\ddag$Department of Electrical and Electronic Engineering, University of Peradeniya, Sri Lanka\\
$^\S$School of Computing and Communications, Lancaster University, United Kingdom \\
Email:  m.a.mohammadi@eng.sku.ac.ir,  b.chalise@csuohio.edu, hakimi@stu.sku.ac.ir, himal@ee.pdn.ac.lk, z.ding@lancaster.ac.uk}}\normalsize

\begin{document}
\maketitle
\thispagestyle{empty}

\begin{abstract}
In this paper, we consider  a non-orthogonal multiple access cognitive radio network, where a full-duplex multi-antenna relay assists transmission from a base station (BS) to a cognitive far user, whereas, at the same time, the BS transmits to a cognitive near user. Our objective is to enlarge the far-near user rate region by maximizing the rate of the near user under a constraint that the rate of the far user is above a certain threshold. To this end, a non-convex joint optimization problem of relay beamforming and the transmit powers at the BS and cognitive relay is solved as a semi-definite relaxation problem, in conjunction with an efficiently solvable line-search approach. For comparisons, we also consider low complexity fixed beamformer design, where the optimum power allocation between the BS and cognitive relay is solved. Our results demonstrate that the proposed joint optimization can significantly reduce the impact of the residual self-interference at the FD relay and inter-user interference in the near user case.
\let\thefootnote\relax\footnotetext{\ThankFour}
\end{abstract}
\section{Introduction}\label{sec:intro}
Non-orthogonal multiple access (NOMA) and full-duplex (FD) communication are foreseen as two independent key technology components of fifth generation (5G) wireless. NOMA exploits power domain to serve multiple users at the same time, frequency and spreading codes~\cite{Saito:VTC2013}. NOMA transmitter sends a superimposed signal with different power levels to the multiple users, where successive interference cancellation (SIC) is utilized to separate superimposed signals at the receiver side and to mitigate the inter-user interference. Therefore, compared to conventional orthogonal multiple access schemes, NOMA can offer a significant improvement in spectrum efficiency~\cite{Zhiguo:COMMG:2017,Zhiguo:CLET:2015}.

On the other hand, FD technology has  been recently received a lot of research interest due to its potential to double the spectrum efficiency and subsequently increase the data rate compared to half-duplex (HD) mode~\cite{Ashutosh:JSAC:2014}. However, the main limitation in FD operation is self-interference (SI) caused by the signal leakage from the transceiver output to the input~\cite{Riihonen:JSP:2011}. Nevertheless, recent progress on FD radio implementations shows great potential for doubling capacity through SI cancellation techniques~\cite{Sabharwal:TWC2012,Khojastepour:Mobicom:2012}.

In the literature, the FD and NOMA combination has been invoked to further enhance the spectral efficiency of the communication systems~\cite{FD-NOMA,Ding:2017:ICC,Caijun:CLET:2016}. The authors of~\cite{FD-NOMA} investigated the resource allocation algorithm design for a FD multicarrier NOMA system, where a FD base station (BS) is simultaneously serving multiple HD downlink and uplink users.  In~\cite{Ding:2017:ICC} a diversity analysis for cooperative FD NOMA systems was provided to prove that the use of the direct link overcomes the lack of diversity for the far user which otherwise serves as a limitation of FD relaying. In~\cite{Caijun:CLET:2016} a dual-user NOMA system has been studied, where a dedicated FD dual-antenna relay assists information transmission to the far user with the weaker channel condition. The proposed FD cooperative NOMA system~\cite{Caijun:CLET:2016} achieves a higher ergodic sum capacity compared to the HD cooperative NOMA counterpart in the low to moderate signal-to-noise (SNR) regimes.

Cognitive radio (CR) is another technology that has received wide attention well over a decade now to improve the spectrum utilization. Under CR paradigm, each cognitive user (CU) is allowed to access the spectrum of the primary users (PUs) as long as the CU meets a certain interference threshold in the primary network (PN)~\cite{Daesik:TWC:2012}. Despite the promise of FD and NOMA for cognitive radio, to the best of our knowledge, current literature has not analyzed such systems. In this work, we investigate a relay assisted cooperative NOMA system at the cognitive network of the CR network, where a cognitive BS communicates with a near and far NOMA CU. The main motivation for the adoption of multiple antennas at the FD relay is that the SI cancellation can be performed in the spatial domain using efficient beamforming design at the cognitive relay~\cite{Riihonen:JSP:2011}. However, beamforming design also influences the achievable NOMA performance of near and far user. We obtain dual-user rate region by designing the receive and transmit beamformers at the relay and allocating optimal power to the BS and cognitive relay. Specifically, the near user rate is maximized by ensuring that the far user rate is above a certain threshold.

The main contributions of this paper are twofold.
\begin{itemize}
\item  A complicated non-convex optimization problem of joint cognitive relay beamforming and power allocation (at the BS and relay) is transformed  to a semi-definite relaxation (SDR) problem consisting of relay transmit beamforming matrix and BS power allocation parameter. The optimum solutions of the joint optimization are obtained by solving the SDR problem in conjunction with a line search over the BS power allocation parameter. The computational complexity of the proposed approach is minimal, since this line search is confined to a finite region and the SDR problem can be solved as a feasibility problem. As compared to traditional HD operation, our new results show that the proposed joint optimization significantly improves the rate region.

\item As a suboptimum approach, we also consider a power allocation problem with fixed beamformer design. We compare the far and near user rate region for the optimum and suboptimum methods to highlight the gains of the proposed optimum design for different system parameters, such as the number of relay transmit/receive antennas, level of the residual SI at the relay and the peak power constraint at both BS and relay.
\end{itemize}

\emph{Notation:} We use bold upper case letters to denote matrices, bold lower case letters to denote vectors. The superscripts  $(\cdot)^{T}$, $(\cdot)^{*}$, $(\cdot)^{\dag}$, and  $(\cdot)^{-1}$  stand for transpose, conjugated, conjugate transpose, and matrix inverse respectively; the Euclidean norm of the vector, the trace, and the expectation are denoted by $\|\cdot\|$, ${\rm tr} (\cdot)$, and ${\tt E}\left\{\cdot\right\}$ respectively; and $\mathcal{CN}(\mu,\sigma^2)$ denotes a circularly symmetric complex Gaussian RV $x$ with mean $\mu$ and variance $\sigma^2$.

\section{System Model}\label{sec:systemmodel}
\begin{figure}
\vspace{-12em}
\centering
\includegraphics[width=80mm, height=100mm]{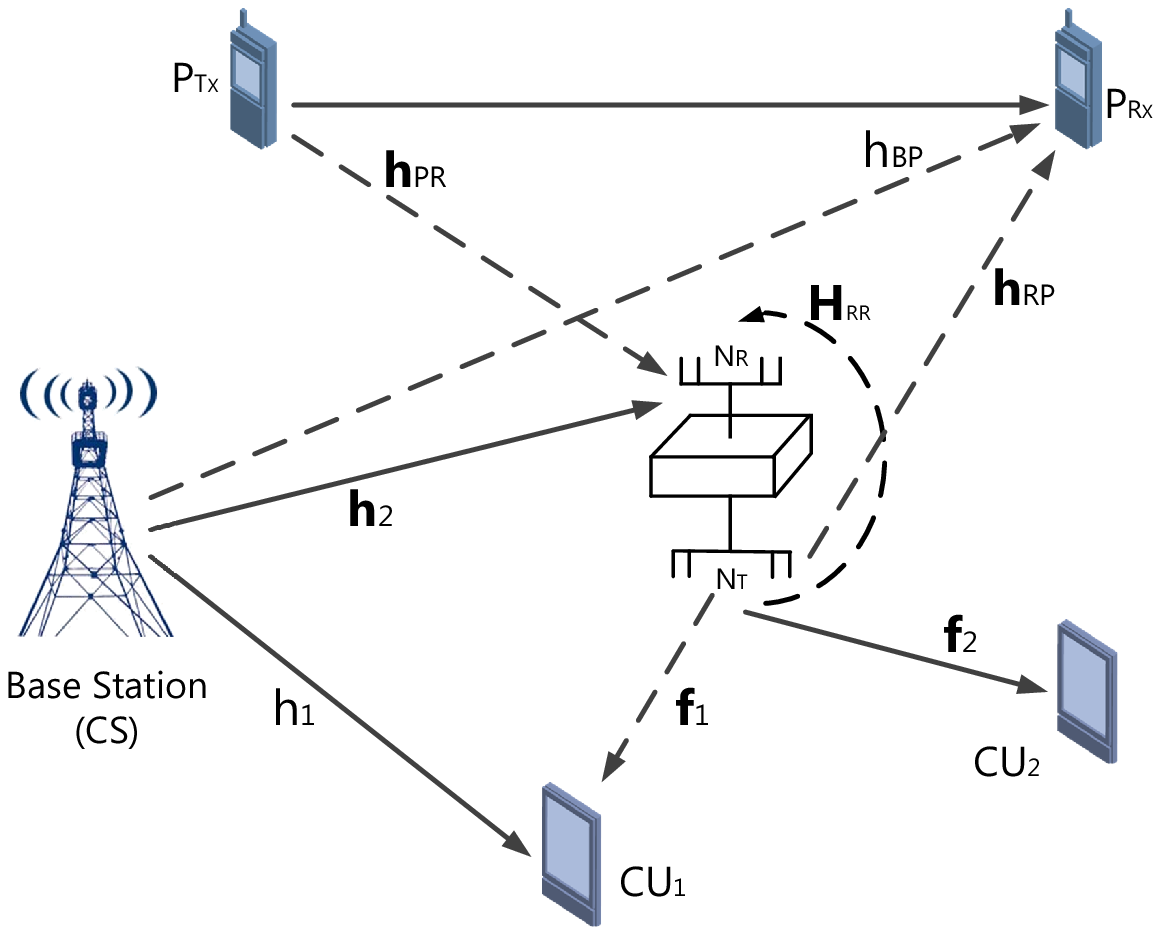}
\vspace{-0.0em}
\caption{Cognitive NOMA system model with FD relaying.}
\label{fig: Cooperative NOMA}
\end{figure}

We consider a FD cognitve relay network as shown in Fig.~\ref{fig: Cooperative NOMA}, where the PUs and CUs share the same spectral band. The cognitive network consists of a BS, a decode-and-forward relay and two CUs, denoted by $\SUu$ and $\SUuu$. The BS communicates with the two CUs, by applying the NOMA concept, where the near user, $\SUu$ directly communicates with the BS, while the far user, $\SUuu$  requires the assistance of the relay. We assume that the BS, $\SUu$ and $\SUuu$ are each equipped with a single antenna~\cite{Caijun:CLET:2016}. To enable FD operation, the cognitive relay is equipped with two sets of antennas, i.e., $\Nrx$  receiveing antennas and $\Ntx$ transmitting antennas.
We assume that no direct link between the BS and $\SUuu$ exists, similar to~\cite{Lee:CLET2015,Caijun:CLET:2016}.

In a spectrum sharing CR system, a CU can share the PU's spectrum as long as the interference inflicted on the primary receiver is below a predetermined maximum tolerable interference level at the PU, $\Ith$~\cite{Daesik:TWC:2012}. Since the BS and relay transmit their signals at the same time using the same spectrum, the primary receiver suffers interference from the BS and cognitive relay simultaneously. Hence, the transmission powers of the BS and cognitive relay must be constrained as~\cite{Daesik:TWC:2012}
\begin{align}\label{eq:power constraint}
\beta_{BP}\Ps|h_{BP}|^2 + \beta_{RP}\Pre|\qh_{RP}^T\qw_t|^2\leq \Ith,
\end{align}
where $\Ps$ and $\Pre$ are the transmission powers of the BS and cognitive relay, $h_{BP}$ and $\qh_{RP}\in\mathcal{C}^{\Ntx\times 1}$ denote the BS-primary receiver channel and the cognitive relay-primary receiver channel respectively, $\beta_{BP}$ and $\beta_{RP}$ model the corresponding path loss effects, and $\qw_t\in\mathcal{C}^{\Ntx\times 1}$ denotes the transmit beamforming vector at the cognitive relay.

Furthermore, similar to the model used in~\cite{JeminLee:TWC:2011, Vincent:TWC:2013,Zhiguo:TVT:2016}, we focus on the coexistence of a long-range primary system and short range CR network. There is a direct link in this set up, however the primary transmitter is far away from the CUs and thus the interference inflicted at the CUs is negligible.

\subsection{Transmission Protocol}
According to the NOMA concept, the BS transmits a combination of intended messages to both CUs as
\vspace{-0.2em}
\begin{align}
s[n]=\sqrt{\Ps a_1}x_1[n]+\sqrt{\Ps a_2}x_2[n],
\end{align}
where $x_i$, $i\in\{1,2\}$ denotes the information symbol intended for $\text{CU}_i$, and $a_i$ denotes the power allocation coefficient,
such that $a_1 + a_2 =1$ and $a_1< a_2$.

The received signal at $\SUu$ can be written as
\vspace{-0.0em}
\begin{align}\label{eq:sig:U1}
y_1[n]\!=\!\sqrt{\beta_{h_1}}h_1s[n]\!+\!\sqrt{\beta_{f_1}\Pre} \qf_1^T \qw_t x_2[n\!-\!\tau]\!+\!n_1[n],
\end{align}
where $h_1$ is the channel between the BS and $\SUu$, $\qf_1\in\mathcal{C}^{\Nrx\times 1}$ denotes the channel between the cognitive relay and $\SUu$, the respective path losses of the BS-$\SUu$ and relay-$\SUu$ links are denoted by $\beta_{h_1}$ and $\beta_{f_1}$, respectively. Further, $\tau$ accounts for the time delay caused by FD relay processing~\cite{Riihonen:JSP:2011}, and $n_1[n]\sim\mathcal{CN}(0,\Snu)$ is the additive white Gaussian noise (AWGN) at $\SUu$.

By invoking~\eqref{eq:sig:U1}, the effective signal-to-interference-plus-noise ratio (SINR) of $\SUuu$ observed at $\SUu$ can be written as
\vspace{-0.0em}
\begin{align}\label{eq:SINR UE2 at UE1}
\gamma_{1,2}&=\frac{\beta_{h_1}\Ps a_2|h_1|^2}{\beta_{h_1}\Ps a_1|h_1|^2 +\beta_{f_1}\Pre|\qf_1^T\qw_t|^2+\Snu}.
\end{align}
It is assumed that symbol $x_{2}[n-\tau]$ is priory known to $\SUu$ and thus $\SUu$ can remove it via interference cancellation~\cite{Caijun:CLET:2016}. However, by considering realistic imperfect interference cancellation wherein $\SUu$ cannot perfectly remove $x_{2}[n-\tau]$, we model $\qf_{1} \sim \mathcal{CN}(0,k_1)$ as the inter-user interference channel where the parameter $k_1$ presents the strength of inter-user interference~\cite{Caijun:CLET:2016}. Specifically, $k_1=0$ implies perfect interference cancellation at $\SUu$. If $\SUu$ cancels the $\SUuu$'s signal, the SINR at $\SUu$ is given by
\vspace{-0.0em}
\begin{align}\label{eq:SINR at UE1}
\gamma_{1}&=\frac{\beta_{h_1}\Ps a_1|h_1|^2}{\beta_{f_1}\Pre|\qf_1^T\qw_t|^2+\Snu}.
\end{align}

The received signal at the cognitive relay can be written  as
\vspace{-0.4em}
\begin{align}
y_R[n]&=\sqrt{\beta_{h_2}}\qw_r^\dag \qh_2s[n]+\sqrt{\Pre}\qw_r^\dag \qH_{RR} \qw_t x_2[n-\tau]\nonumber\\
&+ \sqrt{\beta_{PR}P_U}\qw_r^\dag\qh_{PR} x_P[n]+\qw_r^\dag \qn_R[n],
\end{align}
where $\qw_r\in\mathcal{C}^{\Nrx\times 1}$ is the combining receiver at the cognitive relay, $\qh_2\in\mathcal{C}^{\Nrx\times 1}$ is the channel between the BS and cognitive relay, $P_U$ is transmit power of the primary transmitter, $\qh_{PR}\in\mathcal{C}^{\Nrx\times 1}$ is the channel between the primary transmitter and cognitive relay, $\beta_{h_2}$ and $\beta_{PR}$ model the path loss effect in the BS-cognitive relay channel and primary transmitter-cognitive relay channel, $x_p[n]$ is the primary transmit signal, and $\qn_R[n]$ is the AWGN at the cognitive relay with $\E{\qn_R\qn_R^\dag}=\Snr\qI$. We assume imperfect SI cancellation at the cognitive relay and similar to~\cite{Riihonen:JSP:2011} model the elements of the  $\Nrx\times\Ntx$ residual SI channel $\qH_{RR}$ as independent identically distributed (i.i.d) $\mathcal{CN}(0,\Sap)$ RVs.

The cognitive relay decodes the information intended for $\SUuu$ treating the symbol of $\SUu$ as interference. Hence, the SINR at the cognitive relay can be expressed as
\vspace{-0.4em}
\begin{align}\label{eq:SINR relay}
&\hspace{-0.2em}
\gamma_{R}=\\
&
\hspace{-0.2em}\frac{\beta_{h_2}\Ps a_2|\qw_r^\dag\qh_2|^2}{\beta_{h_2}\Ps a_1|\qw_r^\dag\qh_2|^2 \!+\!\! \Pre|\qw_r^\dag \qH_{RR} \qw_t|^2\!+\!\beta_{PR}P_U|\qw_r^\dag\qh_{PR}|^2 \!\!+\!\Snr}.\nonumber
\end{align}

Moreover, the received signal at $\SUuu$, transmitted by the cognitive relay can be written as
\vspace{-0.4em}
\begin{align}
y_2[n]=\sqrt{\beta_{f_2}\Pre} \qf_2^T\qw_tx_2[n-\tau]+n_2[n],
\end{align}
where $\qf_2\in\mathcal{C}^{\Ntx\times 1}$ denotes the channel between the cognitive relay and $\SUuu$, $\beta_{f_2}$ model the path loss effect of the cognitive relay-$\SUuu$ channel, and $n_2[n]\sim\mathcal{CN}(0,\Snuu)$ denotes the AWGN at the $\SUuu$. Hence, the SNR at $\SUuu$ is given by
\vspace{-0.4em}
\begin{align}\label{eq:gamR2}
\gamma_{R,2}&=\frac{\beta_{f_2}\Pre}{\Snuu }|\qf_2^T\qw_t|^2.
\end{align}

\section{Beamforming Design and Power Allocation}
In this section, we consider joint optimization of receive/transmit beamformers at the cognitive relay and
power allocation at the BS and cognitive relay. We also propose a power allocation scheme between the BS and cognitive relay when fixed beamformers are assumed at the relay. Specifically, we consider maximum ratio transmit (MRT)/maximum ratio combining (MRC) as transmit/receive beamformers, while results for other possible beamformers such as transmit/receive  are left out as future work.

\subsection{Optimum Scheme}
Let us consider the joint design of transmit/receive beamformers and allocation of BS and cognitive relay power such that
achievable rate of $\SUu$ is maximized, while the far user's rate is guaranteed to be above a certain value $\bar{r}$. As such, the optimization problem can be formulated as
\vspace{-0.4em}
\begin{align}\label{eq:opt1}
&\max_{\qw_t, \qw_r,\Ps ,\Pre} \hspace{0.8em} C_1(\qw_t,\Ps,\Pre)  ,\nonumber\\
&\hspace{2em}\text{s.t} \hspace{3.0em} C_2(\qw_t, \qw_r,\Ps,\Pre)\geq \bar{r},\nonumber\\
&\hspace{6em} \beta_{BP}\Ps|h_{BP}|^2 +\beta_{RP} \Pre|\qh_{RP}^T\qw_t|^2\leq \Ith,\nonumber\\
&\hspace{6em} \|\qw_r\| = \|\qw_t\| = 1,\quad~\Ps,\Pre\geq 0,
\end{align}
where~\cite{Caijun:CLET:2016}
\vspace{-0.4em}
\begin{align}
&C_1(\qw_t,\Ps,\Pre)= \log_2\left( 1 + \gamma_1(\qw_t, \Ps, \Pre)\right),\nonumber\\
&C_2(\qw_t, \qw_r,\Ps,\Pre)= \log_2\left( 1 +
\min\left( \gamma_{1,2}(\qw_t, \Ps,\Pre),\right.\right.\nonumber\\
&\hspace{5em}\left.\left.\gamma_{R}(\qw_t,\qw_r, \Ps, \Pre),\gamma_{R,2}(\qw_t, \Pre)\right) \right).
\end{align}
The problem in~\eqref{eq:opt1} can be reformulated as
\vspace{-0.4em}
\begin{align}\label{eq:opt}
&\max_{ \qw_t, \qw_r,\Ps ,\Pre}\hspace{1em} \log_2\left(1\!+\!\frac{\beta_{h_1}\Ps a_1|h_1|^2}{\beta_{f_1}\Pre |\qf_1^T \qw_t|^2+\Snu}\right)  ,\nonumber\\
&\hspace{2em}\text{s.t} \hspace{3em}
\min\left(\gamma_{1,2}(\qw_t, \Ps,\Pre),\gamma_{R}(\qw_t,\qw_r, \Ps, \Pre),\right.\nonumber\\
&\hspace{9em}\left.\gamma_{R,2}(\qw_t, \Pre)\right)\geq \tilde{r},\nonumber\\
&\hspace{6em} \beta_{BP}\Ps|h_{BP}|^2 +\beta_{RP} \Pre|\qh_{RP}^T\qw_t|^2\leq \Ith,\nonumber\\
&\hspace{6em} \|\qw_r\| = \|\qw_t\| = 1,\quad\Ps,\Pre\geq 0,
\end{align}
where $\tilde{r}\triangleq 2^{\bar{r}}-1$. Moreover, the first constraint in~\eqref{eq:opt} can be expressed using the following inequalities:
\vspace{-0.4em}
\begin{subequations}\label{eq:opt2}
\begin{align}
&\beta_{h_1}\Ps a_2|h_1|^2\geq \nonumber\\
&\hspace{1em}\tilde{r}\left(\beta_{h_1}\Ps a_1|h_1|^2+\beta_{f_1}\Pre \qw_t^\dag\qf_1^\ast\qf_1^T\qw_t+\Snu\right) ,\\
& \frac{\beta_{h_2}\Ps a_2\qw_r^\dag\qh_2\qh_2^\dag\qw_r}{\qw_r^\dag\qA\qw_r}\geq\tilde{r},\label{eq:opt2b}\\
& \beta_{f_2}\Pre \qw_t^\dag\qf_2^\ast\qf_2^T\qw_t\geq\Snuu \tilde{r} ,
\end{align}
\end{subequations}
where $\qA = \beta_{h_2}\Ps a_1\qh_2\qh_2^\dag+\Pre \qH_{RR}\qw_t\qw_t^\dag\qH_{RR}^\dag+\beta_{PR}P_U
\qh_{PR}\qh_{PR}^\dag+\Snr \qI$.

By inspecting the optimization problem in~\eqref{eq:opt} we see that only $\gamma_{R}(\qw_t,\qw_r, \Ps, \Pre)$ depends on $\qw_r$. Therefore, it is obvious that the optimum $\qw_r$ is one that maximizes $\gamma_{R}(\qw_t,\qw_r, \Ps, \Pre)$.
Define $g\triangleq\frac{\qw_r^\dag \qh_2\qh_2^\dag\qw_r}{\qw_r^\dag \qA\qw_r}$. The optimum $\qw_r$ is given by
\vspace{-0.4em}
\begin{align}~\label{eq:wropt}
\qw_r=\frac{\qA^{-1}\qh_2}{\|\qA^{-1}\qh_2\|}.
\end{align}
Let $\qB=(\beta_{h_2}\Ps a_1\qh_2\qh_2^\dag+ \beta_{PR}P_U\qh_{PR}\qh_{PR}^\dag+\Snr \qI)$. By substituting $\qw_r$ from~\eqref{eq:wropt} into $g$, we get
\vspace{-0.4em}
\begin{align}\label{eq:opt3}
g&=\qh_2^\dag\left[\Pre \qH_{RR}\qw_t\qw_t^\dag\qH_{RR}^\dag+\qB\right]^{-1}\qh_2\\
& =\qh_2^\dag\qB^{-1}\qh_2-\frac{\left(\qh_2^\dag\qB^{-1}\qH_{RR}\qw_t\qw_t^\dag\qH_{RR}^\dag\qB^{-1}\qh_2\right)\Pre}
{1+\Pre\qw_t^\dag\qH_{RR}^\dag\qB^{-1}\qH_{RR}\qw_t},\nonumber
\end{align}
where we have used the Sherman-Morrison formula, $(\qB + \vu\vv^{\dag})^{-1} = \qB^{-1} - (\qB^{-1}\qu\qv^{\dag}\qB^{-1})/(1 +\qv^{\dag}\qB^{-1}\qu)$ with $\qu=\qv=\sqrt{\Pre}\qH_{RR}\qw_t$.

By substituting~\eqref{eq:opt3} into~\eqref{eq:opt2b}, the optimization problem~\eqref{eq:opt} is reduced w.r.t only $\qw_t$, $\Ps$, and $\Pre$, as follows:
\vspace{-0.4em}
\begin{align}\label{eq:opt4}
&\max_{\qw_t,\Ps ,\Pre}\hspace{0.5em} \log_2\left(1+\frac{\beta_{h_1}\Ps a_1|h_1|^2}{\beta_{f_1}\Pre \qw_t^\dag\qf_1^\ast\qf_1^T\qw_t+\Snu}\right)  ,\nonumber\\
&\hspace{1.5em}\text{s.t}\hspace{1.0em} \hspace{1em} \Pre \qw_t^\dag\qf_1^\ast\qf_1^T\qw_t \leq q_1,\nonumber\\
&\hspace{4.25em} \frac{\left(\qh_2^\dag\qB^{-1}\qH_{RR}\qw_t\qw_t^\dag\qH_{RR}^\dag\qB^{-1}\qh_2\right)\Pre}
{1\!+\!\Pre\qw_t^\dag\qH_{RR}^\dag\qB^{-1}\qH_{RR}\qw_t} \leq q_2,\nonumber\\
&\hspace{4.25em} \Pre \qw_t^\dag\qf_2^\ast\qf_2^T\qw_t\geq q_3\nonumber\\
&\hspace{4.25em} \Pre|\qh_{RP}^T\qw_t|^2\leq q_4,\nonumber\\
&\hspace{4.25em}  \|\qw_t\| = 1, \quad\Ps,\Pre\geq 0,
\end{align}
where
$q_1 \!=\! \left(\beta_{h_1}\Ps a_2|h_1|^2\!-\!\tilde{r}\beta_{h_1}\Ps a_1 |h_1|^2\!-\!\tilde{r}\Snu\right)\!/(\tilde{r}\beta_{f_1})$,
$q_2 = \qh_2^\dag\qB^{-1}\qh_2-\tilde{r}/(\beta_{h_2}\Ps a_2)$, $q_3 = \Snuu \tilde{r}/\beta_{f_2}$, and $q_4=(\Ith- \beta_{BP}\Ps|h_{BP}|^2)/\beta_{RP}$.
From the optimization problem \eqref{eq:opt4}, we see that $\sqrt{\Pre}\qw_t$ can be considered together as a single optimization variable $\bar{\qw_t}$, i.e, $\bar{\qw_t}=\sqrt{\Pre}\qw_t$.
Then \eqref{eq:opt4} reduces to
\vspace{-0.4em}
\begin{subequations}\label{eq:opt5}
\begin{align}
\max_{ \bar{\qw_t},\Ps ,\Pre} &\hspace{1em} \log_2\left(1+\frac{\beta_{h_1}\Ps a_1|h_1|^2}{\beta_{f_1}\bar{\qw_t}^\dag \qf_1^\ast\qf_1^T \bar{\qw_t}+\Snu}\right)  ,\\
\text{s.t} &\hspace{1em} \bar{\qw_t}^\dag\qf_1^\ast\qf_1^T\bar{\qw_t}\leq q_1\,\\
&\hspace{1em} \frac{\qh_2^\dag\qB^{-1}\qH_{RR}\bar{\qw_t}\bar{\qw_t}^\dag\qH_{RR}^\dag\qB^{-1}\qh_2}{1+\bar{\qw_t}^\dag\qH_{RR}^\dag
\qB^{-1}\qH_{RR}\bar{\qw_t}}\leq q_2,\\
&\hspace{1em} \bar{\qw_t}^\dag\qf_2^\ast\qf_2^T\bar{\qw_t}\geq q_3,\label{eq:opt5.4}\\
&\hspace{1em} \bar{\qw_t}^\dag\qh_{RP}^\ast\qh_{RP}^T\bar{\qw_t}\leq q_4,\\
&\hspace{1em} \bar{\qw_t}^\dag\bar{\qw_t}=\Pre, \quad\Ps\geq0.
\end{align}
\end{subequations}
The above problem is a complicated non-convex optimization problem, which to the best of our knowledge, does
not admit a closed-form solution. However, by fixing $\Ps$, optimum $\bar{\qw_t}$ and $\Pre$
can be efficiently obtained. Then the joint optimization over $\Ps$, $\bar{\qw_t}$ and $\Pre$  can be solved by using  one-dimensional search over a finite (also small) region of $\Ps$.

\subsubsection{Optimization over $\bar{\qw_t}$ and $\Pre$ for a given $\Ps$}
Let us find the optimum $\bar{\qw_t}$ and $\Pre$ for a given $\Ps$.
Problem \eqref{eq:opt5} can be alternatively re-expressed as
\vspace{-0.4em}
\begin{subequations}\label{eq:opt6}
\begin{align}
\min_{\bar{\qw_t},\Pre} &\hspace{2em} \bar{\qw_t}^\dag\qf_1^\ast\qf_1^T\bar{\qw_t},\\
\text{s.t} &\hspace{2em} \bar{\qw_t}^\dag\qf_1^\ast\qf_1^T\bar{\qw_t}\leq q_1,\\
&\hspace{2em} \qh_2^\dag\qB^{-1}\qH_{RR}\bar{\qw_t}\bar{\qw_t}^\dag\qH_{RR}^\dag\qB^{-1}\qh_2 \leq\nonumber\\
&\hspace{5em} q_2\left(1+\bar{\qw_t}^\dag\qH_{RR}^\dag\qB^{-1}\qH_{RR}\bar{\qw_t}\right),\label{eq:opt6.3}\\
&\hspace{2em} \bar{\qw_t}^\dag\qf_2^\ast\qf_2^T\bar{\qw_t}\geq q_3,\label{eq:opt6.4}\\
&\hspace{2em} \bar{\qw_t}^\dag\qh_{RP}^\ast\qh_{RP}^T\bar{\qw_t}\leq q_4,\\
&\hspace{2em} \bar{\qw_t}^\dag\bar{\qw_t}=\Pre.
\end{align}
\end{subequations}
The minimum value in \eqref{eq:opt6} will be less than or equal to $q_1$.
Also \eqref{eq:opt6} is a non-convex optimization problem  due to the fact that it is the
minimization of a quadratic function with non-convex quadratic inequality constraints \eqref{eq:opt6.3} and \eqref{eq:opt6.4}.
However, it can be solved using an SDR approach. Introducing $\bar{\qW_t}\triangleq \bar{\qw_t}\bar{\qw_t}^\dag$ and relaxing the rank-one constraint of $\text{rank}({\qW_t})=1$,~\eqref{eq:opt6}  can be expressed as an SDR problem:
\begin{subequations}\label{eq:opt7}
\begin{align}
\min_{\bar{\qW_t},\Pre}  &\hspace{2em}
\trac\left( \bar{\qW_t}\qf_1^\ast\qf_1^T\right),\\
\text{s.t} &\hspace{2em}
\trac\left(\bar{\qW_t}\qf_1^\ast\qf_1^T\right)\leq q_1,\label{eq:opt7.2}\\
&\hspace{2em}
\trac\left( \bar{\qW_t}\qH_{RR}^\dag\qB^{-1}\qh_2\qh_2^\dag\qB^{-1}\qH_{RR}\right)\leq\nonumber\\
&\hspace{3em} q_2\left(1+\trac\left(\bar{\qW}_t\qH_{RR}^\dag\qB^{-1}\qH_{RR}\right)\right),\\
&\hspace{2em} \trac\left(\bar{\qW_t}\qf_2^\ast\qf_2^T\right)\geq q_3,\\
&\hspace{2em} \trac\left(\bar{\qW_t}\qh_{RP}^\ast\qh_{RP}^T\right)\leq q_4,\\
&\hspace{2em} \trac\left(\bar{\qW_t}\right)=\Pre,
\bar{\qW_t}\geq0.
\end{align}
\end{subequations}

This problem can be solved using CVX software~\cite{Boyd:CVX}. Problem~\eqref{eq:opt7} can be solved without~\eqref{eq:opt7.2}, however if the optimum $\bar{\qW_t}$ does not satisfy~\eqref{eq:opt7.2}, the problem is infeasible. Moreover, $P_S$ and ${\tilde r}$ have to be chosen such that~\eqref{eq:opt7} is feasible. To this end, initialize $P_S$ so that $q_1$, $q_2$ and $q_4$ are positive. From $q_1\geq0$, we get
\vspace{-0.4em}
\begin{align}\label{eq:con:Ps:q1}
\Ps\geq \frac{\tilde{r}\Snu}
{\beta_{h_1}a_2|h_1|^2\!-\!\tilde{r}\beta_{h_1}a_1 |h_1|^2},
\end{align}
which is established when $\beta_{h_1}a_2|h_1|^2\!-\!\tilde{r}\beta_{h_1}a_1 |h_1|^2 >0$, or equivalently when $\tilde{r} < \frac{a_2}{a_1}$.
Moreover, from $q_2\geq0$ we have
\vspace{-0.3em}
\begin{align}
q_2 &= \qh_2^\dag\qB^{-1}\qh_2-\frac{\tilde{r}}{\beta_{h_2}\Ps a_2}\geq0.
\end{align}
Recall that $\qB=(\beta_{h_2}\Ps a_1\qh_2\qh_2^\dag+ \beta_{PR}P_U\qh_{PR}\qh_{PR}^\dag+\Snr \qI)$ is a function of $P_S$. Define $\qE \triangleq \left(\beta_{PR}P_U\qh_{PR}\qh_{PR}^\dag+\Snr\qI\right)$ and  $u\triangleq \qh_2^\dag\left(\qE+\beta_{h_2}\Ps a_1\qh_2\qh_2^\dag\right)^{-1}\qh_2$. Then, using Sherman-Morrison formula we get
\begin{algorithm}~\label{alg:optimization}
 \caption{ The proposed optimization scheme}
 \begin{algorithmic}

  \renewcommand{\algorithmicrequire}{\textbf{Step 1:} }
   \REQUIRE Define a fine grid of $\Ps$, where $\Ps \in \left[v, \frac{\Ith}{\beta_{BP}|h_{BP}|^2}\right]$, in steps of $\delta \Ps$. Set  $\Ps=\frac{\Ith}{\beta_{BP}|h_{BP}|^2}$.

 \renewcommand{\algorithmicrequire}{\textbf{Step 2:} }
 \REQUIRE Solve \eqref{eq:opt7}.

\renewcommand{\algorithmicrequire}{\textbf{Step 3:} }
\REQUIRE  If feasible, stop and output $\Ps$, $\Pre$, and $\bar{\qW_t}$.

\renewcommand{\algorithmicrequire}{\textbf{Step 4:} }
\REQUIRE If not, go to {\textbf{Step 2} } with the decrement of $\delta \Ps$.

\end{algorithmic}
 \end{algorithm}
\begin{align}
u&=\qh_2^\dag \left[\qE^{-1}-\frac{\qE^{-1}\qh_2\qh_2^\dag\beta_{h_2}\Ps a_1\qE^{-1}}{1+\beta_{h_2}\Ps a_1\qh_2^\dag\qE^{-1}\qh_2}\right]\qh_2\nonumber\\
&=\frac{\qh_2^\dag\qE^{-1}\qh_2}{1+\beta_{h_2}\Ps a_1\left(\qh_2^\dag\qE^{-1}\qh_2\right)}.
\end{align}

Next, by substituting $u$ into $q_2$, it is clear that $q_2\geq 0$ if
\vspace{-0.4em}
\begin{align}\label{eq:con:Ps:q2}
\Ps\geq \frac{{\tilde r}}{(a_2-{\tilde r} a_1) \beta_{h_2} \qh_2^\dag\qE^{-1}\qh_2}.
\end{align}
Finally, from $q_4\geq 0$, we get
\vspace{-0.4em}
\begin{align}\label{eq:con:Ps:q4}
\Ps \leq \frac{\Ith}{\beta_{BP}|h_{BP}|^2}.
\end{align}
From the conditions~\eqref{eq:con:Ps:q1}, ~\eqref{eq:con:Ps:q2}, and~\eqref{eq:con:Ps:q4} on $P_S$, it is clear that the optimization problem \eqref{eq:opt7} is feasible if
\vspace{-0.4em}
\begin{eqnarray}
\label{eq:mainCond1}
& v \triangleq \max\limits_{}\left \{  \frac{{\tilde r}}{(a_2-{\tilde r} a_1) \beta_{h_2} \qh_2^\dag\qE^{-1}\qh_2}, \frac{\tilde{r}\Snu}
{\beta_{h_1}(a_2-\tilde{r}a_1) |h_1|^2}  \right\} \nonumber\\
& \hspace{1em} \leq  \frac{\Ith}{\beta_{BP}|h_{BP}|^2},
\end{eqnarray}
which also means that $\Ps \in \left[v, \frac{\Ith}{\beta_{BP}|h_{BP}|^2}\right]$.

\subsubsection{Joint Optimization of $\Ps$, $\Pre$, and $\bar{\qw_t}$}
The joint optimization problem~\eqref{eq:opt5} can be solved by solving the SDR problem (\ref{eq:opt7}) for different values of $\Ps$ (i.e., performing line search over $\Ps$), where $\Ps \in \left[v, \frac{\Ith}{\beta_{BP}|h_{BP}|^2}\right]$,  and taking those values of $\bar{\qw_t}$ and $\Ps$ that maximize the objective function in \eqref{eq:opt5}. However, it is clear that this objective function monotonically increases with $\Ps$. This means that the optimum $\Ps$ is its largest value, for which the problem
(\ref{eq:opt7}) is feasible. As such, starting with $\Ps =\frac{\Ith}{\beta_{BP}|h_{BP}|^2}$,  the joint optimization problem~\eqref{eq:opt5}  can be solved by solving  (\ref{eq:opt7}) until it turns to be feasible. This leads to an iterative approach, which is outlined in  {\bf Algorithm 1}.

We end this subsection with the following remark.\\
{\it Remark:} Applying Shapiro-Barvinok-Pataki rank reduction result, it can be shown that rank-one optimum solution of  $\bar{\qW_t}$ exists for the  SDR problem (\ref{eq:opt7}) \cite{ChaliseMaTSP2013}. This allows us to recover $\bar{\qw_t}$ from $\bar{\qW_t}$, without any loss of optimality. Moreover, if the optimum $\bar{\qW_t}$ is rank-one,  $\bar{\qw_t}$ is given by $\bar{\qw_t}=\lambda_{\rm max}{\bf v}_{\rm max}$, where ${\bf v}_{\rm max}$ is the eigenvector corresponding to the largest eigenvalue, $\lambda_{\rm max}$, of  $\bar{\qW_t}$. Due to these reasons, relaxation in (\ref{eq:opt7}) is optimum. As such, the proposed iterative algorithm finds the optimum solutions of $\Ps$, $\Pre$, and $\qw_t$. Moreover, the computational cost of implementing the algorithm is minimal, since the search region of $\Ps$ is found to be finite and the algorithm can be stopped as soon as the optimization  \eqref{eq:opt7} is feasible.

\subsection{Power Allocation for fixed Beamforming Design}
In this subsection, we further investigate the power allocation problem by considering fixed $\qw_t$ and $\qw_r$. The motivation for considering a fixed choice for $\qw_t$ and $\qw_r$  is as follows: Fixed beamformers constitute to low complex implementation. For example, MRT/MRC beamformers are suitable for low complexity FD systems as they do not need to estimate the SI channel. Moreover, MRT/MRC beamformers are preferred for HD operation and hence it is interesting to characterize the achievable performance in the FD case.

For a given $\qw_t$ and $\qw_r$, the optimization problem~\eqref{eq:opt1} is expressed as
\begin{align}\label{eq:opt1:fixed}
&\max_{\Ps ,\Pre} \hspace{0.8em} C_1(\qw_t,\Ps,\Pre)  ,\nonumber\\
&\hspace{2em}\text{s.t} \hspace{3.0em} C_2(\qw_t, \qw_r,\Ps,\Pre)\geq \bar{r},\nonumber\\
&\hspace{6em} \beta_{BP}\Ps|h_{BP}|^2 +\beta_{RP} \Pre|\qh_{RP}^T\qw_t|^2\leq \Ith,\nonumber\\
&\hspace{6em} \Ps,\Pre\geq 0.
\end{align}

The problem~\eqref{eq:opt1:fixed} can be reformulated as
\begin{align}\label{eq:PA:TZF}
&\max_{\Ps ,\Pre}\hspace{1em}
\log_2\left(1\!+\!\frac{\Ps \beta_{h_1}a_1|h_1|^2}{\Pre \beta_{f_1}|\qf_1^T \qw_t|^2+\Snu}\right)  ,\nonumber\\
&\hspace{1em}\text{s.t}
\hspace{2em}
\frac{\Ps \beta_{h_1}a_2|h_1|^2}{\Ps \beta_{h_1}a_1|h_1|^2 +\Pre\beta_{f_1}|\qf_1^T \qw_t|^2+\Snu}\geq \tilde{r},\nonumber\\
&\hspace{4em}\frac{\Ps\beta_{h_2} a_2|\qw_r^{\dag}\qh_2|^2}{\Ps \beta_{h_2} a_1|\qw_r^{\dag}\qh_2|^2 \!
+\Pre|\qw_r^\dag\qH_{RR}\qw_t|^2
+
\mu_1
}\geq \tilde{r},\nonumber\\
&\hspace{4em} \Pre\frac{\beta_{f_2}}{\Snuu }|\qf_2^T\qw_t|^2\geq \tilde{r},\nonumber\\
&\hspace{4em} \Ps \beta_{BP}|h_{BP}|^2 + \Pre\beta_{RP}|\qh_{RP}^T\qw_t|^2\leq I_{th},\nonumber\\
&\hspace{4em}\Ps,\Pre\geq 0,
\end{align}
where $\mu_1\triangleq\beta_{PR}P_U|\qw_r^{\dag}\qh_{PR}|^2 \!+\!\Snr$.
Accordingly, when $\tilde{r}< \frac{a_2}{a_1}$, the problem~\eqref{eq:PA:TZF} can be expressed as
\begin{align}\label{eq:PA:TZF2}
&\max_{\Ps ,\Pre}\hspace{1em} \frac{\Ps \beta_{h_1}a_1|h_1|^2}{\Pre\beta_{f_1}|\qf_1^T \qw_t|^2+\Snu}  ,\nonumber\\
&\hspace{1em}\text{s.t}
\hspace{2em} \Ps  \geq \frac{\Pre \beta_{f_1}|\qf_1^T \qw_t|^2 + \Snu}{\frac{\beta_{h_1}a_2|h_1|^2}{\tilde{r}}-\beta_{h_1}a_1|h_1|^2},\nonumber\\
&\hspace{4em} \Ps  \leq \frac{I_{th}-\Pre \beta_{RP}|\qh_{RP}^T\qw_t|^2 }{\beta_{BP}|h_{BP}|^2},\nonumber\\
&\hspace{4em} \Ps  \geq \frac{\Pre|\qw_r^\dag\qH_{RR}\qw_t|^2+\mu_1}{\frac{\beta_{h_2} a_2|\qw_r^{\dag}\qh_2|^2}{\tilde{r}}-\beta_{h_2} a_1|\qw_r^{\dag}\qh_2|^2},\nonumber\\
&\hspace{4em} \Pre \geq \frac{\Snuu\tilde{r}}{{\beta_{f_2}}|\qf_2^T\qw_t|^2},,\nonumber\\
&\hspace{4em}\Pre \leq \frac{I_{th}}{\beta_{RP}|\qh_{RP}^T\qw_t|^2}.
\end{align}
Note that a feasible solution of $\Pre$ exists in  (\ref{eq:PA:TZF2}) if $\frac{\Snuu\tilde{r}}{{\beta_{f_2}}|\qf_2^T\qw_t|^2} \leq  \frac{I_{th}}{\beta_{RP}|\qh_{RP}^T\qw_t|^2}$. Similarly, a feasible solution of $\Ps$ exists if
 ${\tilde v}(\Pre) \leq  \frac{I_{th}-\Pre \beta_{RP}|\qh_{RP}^T\qw_t|^2 }{\beta_{BP}|h_{BP}|^2}$, where
\begin{align}
{\tilde v}(\Pre) &\triangleq \max\biggl\{   \frac{\Pre \beta_{f_1}|\qf_1^T \qw_t|^2 + \Snu}{ \beta_{h_1}|h_1|^2\left( \frac{a_2}{\tilde{r}}-a_1\right)},  \nonumber\\
&\hspace{5em}\frac{\Pre|\qw_r^\dag\qH_{RR}\qw_t|^2+\mu_1}{\beta_{h_2}|\qw_r^{\dag}\qh_2|^2 \left( \frac{a_2}{\tilde{r}}-a_1\right)}\biggr\}.
\end{align}
This also means that $\Ps \in \left[ {\tilde v}(\Pre) ,    \frac{I_{th}-\Pre \beta_{RP}|\qh_{RP}^T\qw_t|^2 }{\beta_{BP}|h_{BP}|^2} \right]$.
On the other hand, the objective function in (\ref{eq:PA:TZF2}) is maximized with the minimum value of $\Pre$ and the maximum value of $\Ps$.  Clearly, the minimum value of  $\Pre$, i.e., $\Pre = \frac{\Snuu\tilde{r}}{{\beta_{f_2}}|\qf_2^T\qw_t|^2}$ provides the largest region for feasible solutions of $\Ps$.  Any other $\Pre >\frac{\Snuu\tilde{r}}{{\beta_{f_2}}|\qf_2^T\qw_t|^2}$ contracts this region.  For $\Pre = \frac{\Snuu\tilde{r}}{{\beta_{f_2}}|\qf_2^T\qw_t|^2}$, the maximum possible value of $\Ps$ is given by
$\Ps=(I_{th}-\frac{\Snuu\tilde{r}}{{\beta_{f_2}}|\qf_2^T\qw_t|^2} \beta_{RP}|\qh_{RP}^T\qw_t|^2 )/\beta_{BP}|h_{BP}|^2$.
 Consequently,  the optimum solutions of $\Pre$ and $\Ps$ are given by
\begin{eqnarray}
\Pre &=&\frac{\Snuu\tilde{r}}{{\beta_{f_2}}|\qf_2^T\qw_t|^2}, \nonumber\\
\Ps&=&\frac{I_{th}-\frac{\Snuu\tilde{r}}{{\beta_{f_2}}|\qf_2^T\qw_t|^2} \beta_{RP}|\qh_{RP}^T\qw_t|^2 }{\beta_{BP}|h_{BP}|^2}.
\end{eqnarray}

\begin{figure}
\centering
\includegraphics[width=80mm, height=62mm]{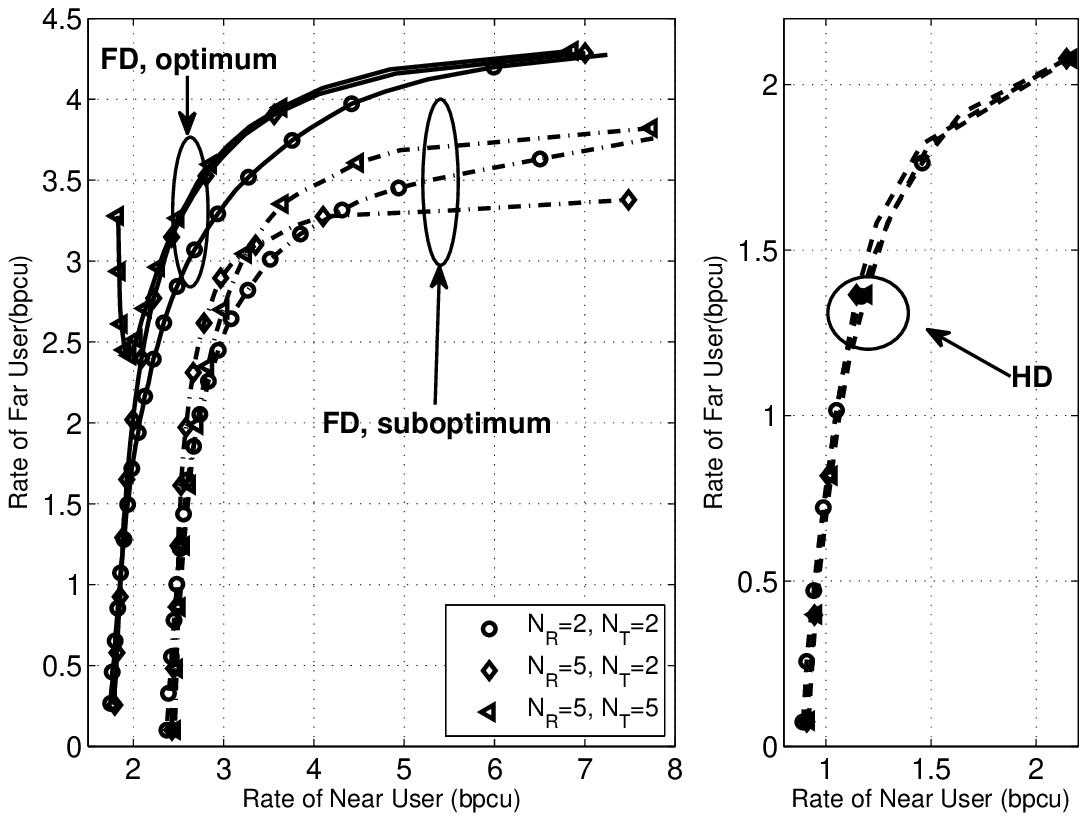}
\caption{Rate-region of the optimum and suboptimum schemes for different antenna configurations at the relay. ($\Ith=15$ dBW, $k_1=0.01$)}
\label{fig: rate region Antconf}
\vspace{0em}
\end{figure}
\section{Numerical and Simulation Results}\label{sec:num}
In this section, we present numerical results to evaluate the rate region of $\SUu$ and $\SUuu$ due to the optimum and fixed beamformer designs. Without loss of generality, the noise variance are set to $1$ dBW, $P_U=10$ dBW, $a_1=0.05$ and $a_2=0.95$. We also adopt the same channel parameters as in~\cite{Caijun:CLET:2016}. Hence, we set $\beta_{BP} = \beta_{RP}=  \beta_{h_2}=  \beta_{f_2} =0.5$ and $\beta_{h_1} =1$. We also show results for the HD mode where comparisons between FD and HD were performed under the ``RF-chain preserved" condition~\cite{Khojastepour:Mobicom:2012}.


Fig.~\ref{fig: rate region Antconf} shows the rate-region of the optimum and suboptimum schemes for different antenna configurations at the relay. The rate-region of the HD mode is shown with the BS and cognitive
relay transmit powers  set as $\frac{\Ith}{\beta_{BP}|h_{BP}|^2}$ and $\frac{\Ith}{\beta_{RP}|h_{RP}^T\qw_t|^2}$, respectively.
Moreover, MRC/MRT processing is shown as an example of  fixed beamforming design case. From the figure, we can observe that the rate of both near and far users with optimum scheme is improved when the number of transmit or receive antenna is increased. Specifically, this increase is more pronounced when the number of the receive antennas increases. However, in case of fixed beamforming design, when only $\Nrx$ is increased from 2 to 5, the rate of the far user decreases, at higher achievable rates for the near user. This is quite intuitive since to achieve higher rate at the near user, the relay transmit power  must be increased.
However, an increase in relay transmit power results in strong SI at the relay input which will degrade the performance.  With more transmit antennas at the relay, transmit power of the relay can be controlled more precisely and hence both near and far user rates are increased. Moreover, comparing FD and HD modes of operation, we see that the FD mode with the optimum and suboptimum schemes provides superior rates for both the near and far users.

\begin{figure}
\vspace{0em}
\centering
\includegraphics[width=80mm, height=57mm]{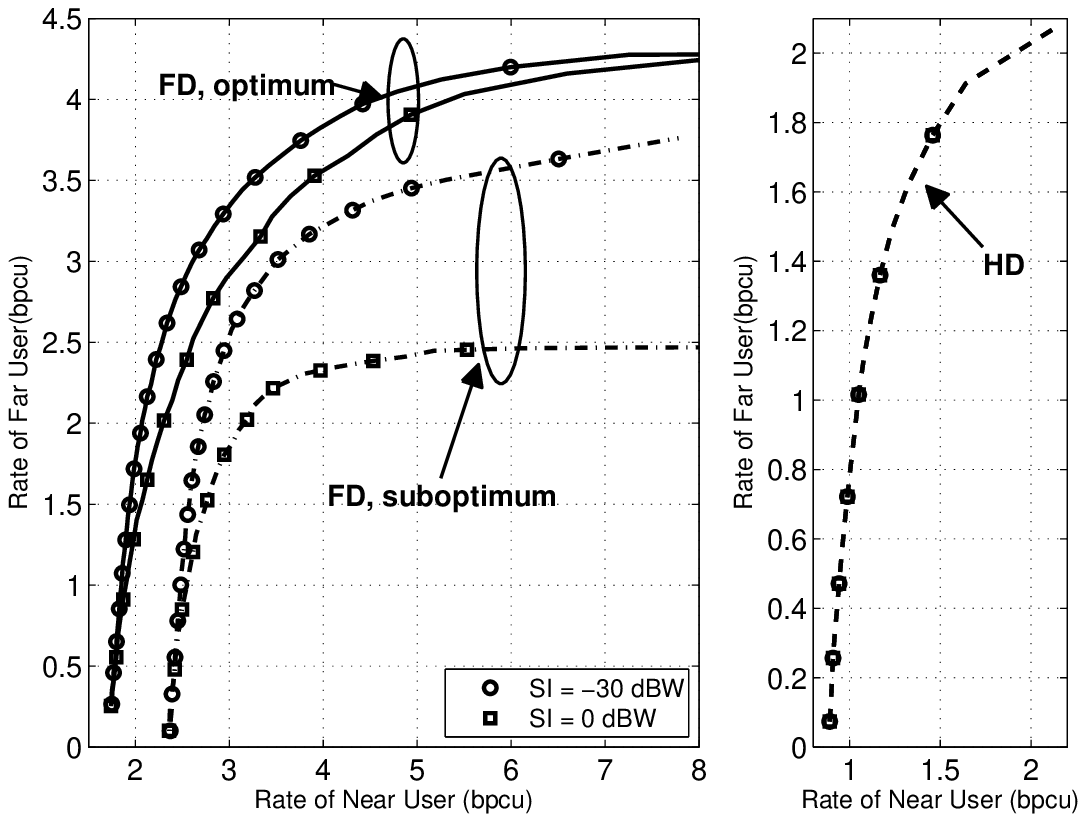}
\vspace{-0.5em}
\caption{Rate-region of the optimum and suboptimum schemes for different levels of SI strength. ($\Ntx=\Nrx=2$, $\Ith=15$ dBW, $k_1=0.01$)}
\label{fig: rate region SI}
\vspace{-0.0em}
\end{figure}

Fig.~\ref{fig: rate region SI} shows the impact of the residual SI on the rate region of the optimum and suboptimum schemes. As expected, the rate of the far user, with the optimum and suboptimum schemes, degrades when the residual SI becomes stronger, while the rate of the far user with HD mode remains the same regardless of the SI power level. More specifically  it can be seen that, the decrease of the far user's rate associated with the  optimum scheme is strictly smaller than that of the suboptimum scheme which indicates that our proposed joint beamforming design and power allocation scheme at the FD relay could significantly suppress the residual SI and consequently can improve the rate of the far user.

\begin{figure}
\vspace{0em}
\centering
\includegraphics[width=80mm, height=57mm]{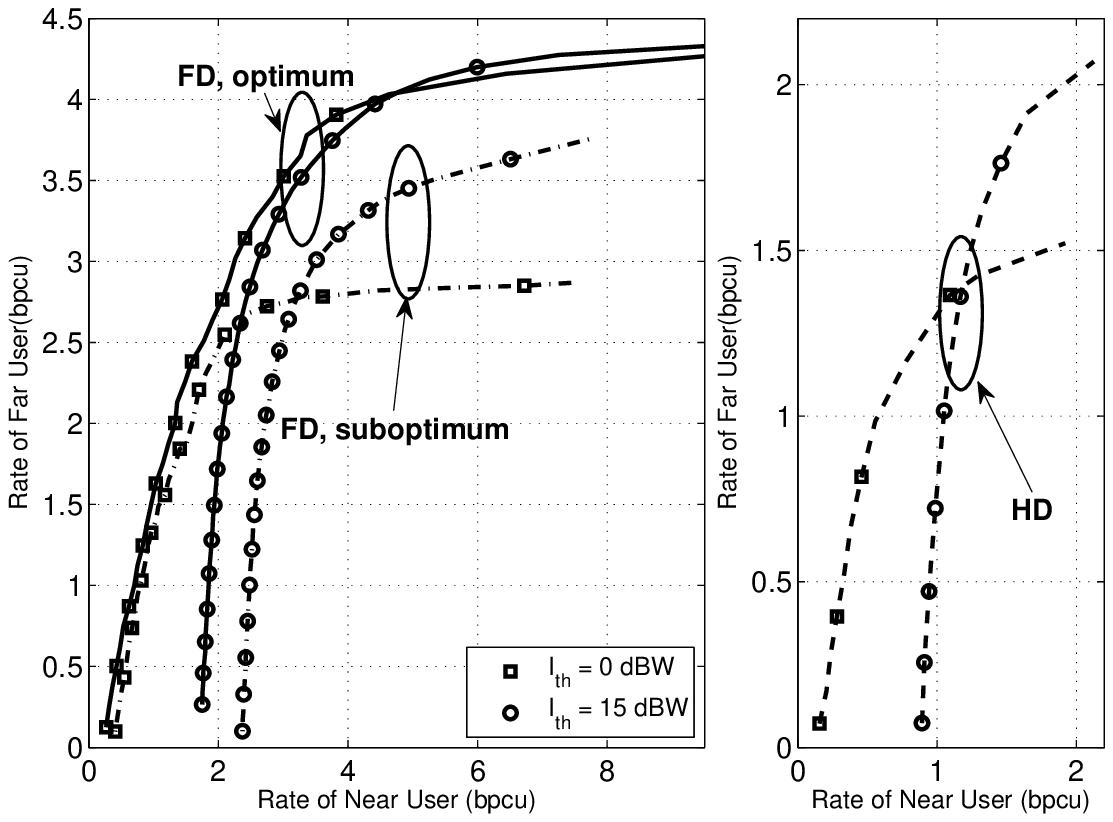}
\vspace{-0.5em}
\caption{Rate-region of the optimum and suboptimum schemes for different values of $\Ith$. ($\Ntx=\Nrx=2$, $\Sap=-30$ dBW, $k_1=0.01$)}
\label{fig: rate region Ith}
\vspace{-0.0em}
\end{figure}

Fig.~\ref{fig: rate region Ith} compares the rate region of the optimum and suboptimum schemes for different levels of $\Ith$ at the primary receiver. It can be readily observed that the gap between the achievable rate of the far user for optimum and suboptimum schemes increases when $\Ith$ decreases. This is because that the BS and relay transmit power are decreased and hence the rate of the far user is decreased. On the other hand, employing the joint beamforming design and power allocation improves the far user rate significantly and hence there is a slight gap between the far user's rate with low and moderate values of $\Ith$.

\section{Conclusion}\label{sec:conclusion}
In this paper, we have investigated the rate region of near and far users in a FD relay assisted NOMA cognitive radio network. An optimum scheme was proposed to maximize the rate of the near user through the joint receive and transmit beamforming and power allocation design at the cognitive relay, by ensuring that the rate of the far user is above a certain threshold. In addition we considered suboptimal design in which a power allocation solution was derived for any fixed receive and transmit beamforming design at the FD relay. Our result indicate that FD relaying with proposed optimum and suboptimum schemes can substantially boost both near and far user rates as compared to the HD mode.
\bibliographystyle{IEEEtran}

\end{document}